\documentclass{article}
\topskip=\baselineskip
\let\al=\alpha
\let\bt=\beta
\let\gm=\gamma
\let\dl=\delta
\let\ep=\epsilon

\let\sg=\sigma
\let\la=\langle
\let\ra=\rangle
\let\pa=\partial
\let\e=\emph

\let\ct=\cite

\let\bv=\mathbf
\let\mr=\mathrm
\let\dt=\cdot
\let\del=\nabla

\let\q=\widehat
\let\Q=\overbrace
\let\h=\hbar
\let\rta=\rightarrow
\let\dy=\displaystyle

\let\hl=\hfill
\newcommand{\m}{\mbox}
\newcommand{\ol}[1]{\makebox[\textwidth][s]{#1}}
\newcommand{\id}{\mathrm{I}}
\newcommand{\eqdf}{\stackrel{\mathrm{def}}{=}}
\newcommand{\hf}{\ensuremath{{\scriptstyle\frac{1}{2}}}}
\newcommand{\hfs}{\ensuremath{{\scriptscriptstyle\frac{1}{2}}}}

\newcommand{\be}{\begin{equation}}
\newcommand{\ee}{\end{equation}}
\newcommand{\dd}[3]{\\ \m{}\\ \ol{\m{#1}\hl\m{${\dy #2}$}\hl\m{#3}}\\ \m{}\\}
\newcommand{\re}[2]{\dd{}{#1}{(#2)}}

\newcommand{\ba}{\begin{array}}
\newcommand{\ea}{\end{array}}
\newcommand{\bea}{\begin{eqnarray}}
\newcommand{\eea}{\end{eqnarray}}
\newcommand{\beas}{\begin{eqnarray*}}
\newcommand{\eeas}{\end{eqnarray*}}

\newcommand{\qH}{\q{H}}
\newcommand{\Hfr}{H_{\mr{free}}}
\newcommand{\qHfr}{\q{H}_{\mr{free}}}
\newcommand{\Hpl}{H_{\mr{Pauli}}}
\newcommand{\Hnz}{H^{\mr{(NR)}}_{\mr{EM;0}}}
\newcommand{\Hrz}{H^{\mr{(REL)}}_{\mr{EM;0}}}
\newcommand{\Hnh}{H^{\mr{(NR)}}_{\mr{EM;}\hfs}}
\newcommand{\Hrh}{H^{\mr{(REL)}}_{\mr{EM;}\hfs}}

\newcommand{\vx}{\bv{x}}
\newcommand{\vv}{\bv{v}}
\newcommand{\vp}{\bv{p}}
\newcommand{\vP}{\bv{P}}

\newcommand{\vA}{\bv{A}}

\newcommand{\vL}{\bv{L}}

\newcommand{\vz}{\bv{0}}

\newcommand{\qvx}{\q{\vx}}
\newcommand{\qvp}{\q{\vp}}
\newcommand{\qvP}{\q{\vP}}
\newcommand{\qvL}{\q{\vL}}

\newcommand{\qp}{\q{p}}

\newcommand{\qP}{\q{P}}
\title{Sound relativistic quantum mechanics for a \\
        strictly solitary nonzero-mass particle, \\
          and its quantum-field reverberations}
\author{Steven Kenneth Kauffmann \\
        American Physical Society Senior Life Member}
\date{43 Bedok Road \\
      {\#}01-11 \\
      Country Park Condominium \\
      Singapore 469564 \\
      Handphone: +65 9370 6583 \\
      \m{} \\
      and \\
      \m{} \\
      Unit 802, Reflection on the Sea \\
      120 Marine Parade \\
      Coolangatta QLD 4225 \\
      Australia \\
      Tel/FAX: +61 7 5536 7235 \\
      Mobile:  +61 4 0567 9058 \\
      \m{} \\
      Email: SKKauffmann@gmail.com}
\begin{document}
\maketitle
\begin{abstract}
It is generally acknowledged that neither the Klein-Gordon equation nor the
Dirac Hamiltonian can produce sound solitary-particle relativistic quantum
mechanics due to the ill effects of their negative-energy solutions; instead
their field-quantized wavefunctions are reinterpreted as dealing with particle
and antiparticle simultaneously---despite the clear physical distinguishability
of antiparticle from particle and the empirically known slight breaking of the
underlying CP invariance.  The natural square-root Hamiltonian of the free
relativistic solitary particle is iterated to obtain the Klein-Gordon equation
and linearized to obtain the Dirac Hamiltonian, steps that have calculational
but not physical motivation, and which generate the above-mentioned problematic
negative-energy solutions as extraneous artifacts.  Since the natural square-%
root Hamiltonian for the free relativistic solitary particle contrariwise
produces physically unexceptionable quantum mechanics, this article focuses on
extending that Hamiltonian to describe a solitary particle (of either spin~0 or
spin~$\hf$) in relativistic interaction with an external electromagnetic
field.  That is achieved by use of Lorentz-covariant solitary-particle four-%
momentum techniques together with the assumption that well-known nonrelativistic
dynamics applies in the particle's rest frame.  Lorentz-invariant solitary-%
particle actions, whose formal Hamiltonization is an equivalent alternative
approach, are as well explicitly displayed.  It is proposed that two separate
solitary-particle wavefunctions, one for a particle and the other for its
antiparticle, be independently quantized in lieu of ``reinterpreting''
negative-energy solutions---which indeed don't even afflict proper solitary
particles.
\end{abstract}

\subsection*{Introduction}

Motivated by certain considerations of perceived calculational ease
\e{rather than by any compelling physical argument}~\ct{B-D},
Klein, Gordon and Schr\"{o}dinger \e{iterated} the \e{natural}
Schr\"{o}dinger equation for a free relativistic solitary
nonzero-mass particle,
\re{
i\h\partial |\psi\rangle /\partial t = \sqrt{m^2c^4 + |c\qvp |^2}
 |\psi\rangle,
}{1}
to become,
\[ -\h^2\partial^2|\psi\rangle /\partial t^2 =
(\sqrt{m^2c^4 + |c\qvp |^2}\:)^2 |\psi\rangle =
(m^2c^4 + |c\qvp |^2)|\psi\rangle.\]
We see that this adds to each to each stationary eigensolution
$e^{-i\sqrt{m^2c^4 + |c\vp |^2}t/\h}|\vp\rangle$ of the above
relativistic free solitary-particle Schr\"{o}dinger equation
an \e{extraneous} negative-energy partner solution
$e^{+i\sqrt{m^2c^4 + |c\vp |^2}t/\h}|\vp\rangle$.  These
extraneous \e{negative} ``free solitary-particle'' energies,
$-\sqrt{m^2c^4 + |c\vp |^2}$, do \e{not} correspond to \e{anything}
that exists in the \e{classical} dynamics of a free relativistic solitary
particle, and by their negatively \e{unbounded} character threaten
to spawn unstable runaway phenomena should the Klein-Gordon equation be
sufficiently perturbed (the Klein paradox)~\ct{B-D}.  Since the
Klein-Gordon equation \e{lacks} a corresponding Hamiltonian, it turns
out, as is easily verified, that the \e{two} solutions of the \e{same
momentum} $\vp$ which have \e{opposite-sign} energies, i.e.,
$\pm\sqrt{m^2c^4 + |c\vp |^2}$, \e{fail} to be \e{orthogonal} to each
other, which \e{violates a key property} of orthodox quantum mechanics.
\e{Without this property} the probablity interpretation of quantum
mechanics \e{cannot be sustained}, and the Klein-Gordon equation is
unsurprisingly diseased in that regard, yielding, inter alia,
\e{negative} probabilities~\ct{B-D}.

This probability disease prompted Dirac to try to replace the Klein-%
Gordon equation with a \e{Hamiltonian}, specifically a \e{linearization}
of the natural relativistic free-particle Hamiltonian $\sqrt{m^2c^4 +
|c\qvp|^2}$ of Eq.~(1) that has the postulated form $\q{H}_D = \al_0mc^2
+ \vec\al\dt\qvp c$, where imposition on the Hermitian
matrices $(\al_0,\al_1,\al_2,\al_3)$
of the anticommutation relations, $\al_r\al_s + \al_s\al_r = 2\dl_{rs}$,
$r,s = 0, 1, 2, 3$, ensures that~\ct{B-D},
\[\q{H}_D^2 = m^2c^4 + |c\qvp |^2 = (\sqrt{m^2c^4 + |c\qvp |^2}\:)^2.\]
Dirac's motivation for \e{linearizing} the natural relativistic free-%
particle Hamiltonian $\sqrt{m^2c^4 + |c\qvp|^2}$ was \e{again} one of
perceived calculational ease \e{rather than any compelling physical
argument}.  The eigenenergies of Dirac's linearized $\q{H}_D$ turn out to
\e{include} all the extraneous \e{negative} energies which are such a
vexing feature of the Klein-Gordon equation's solutions in the context of
a free solitary particle.  Technically, this is a consequence of the fact
that, as a matrix, $\q{H}_D$ is traceless because each of
the four matrices $\al_r$, $r = 0, 1, 2, 3$, is traceless, as can be
demonstrated by using their anticommutation relations~\ct{B-D}.
While the negative-energy eigenstates of $\q{H}_D$ \e{are}
properly \e{orthogonal} to their positive-energy counterparts, the
\e{other} inherent issues which the presence of these negative-energy
solutions raise in the context of a free solitary particle, such as the
classical limit and the Klein paradox remain unresolved~\ct{B-D}.  In
addition, straightforward calculation of the free solitary-particle
velocity and consequent speed using $\q{H}_D$ and the Heisenberg equa%
tion of motion reveals apparent incompatibility of this Hamiltonian with
special relativity (i.e., a \e{universal} free-particle speed of
$\sqrt{3}c$) when it is interpreted as a \e{strictly solitary-particle}
Hamiltonian.  (On this basis one \e{also} finds that $\q{H}_D$ implies
an egregious violation of Newton's first law of motion for a free solitary
particle, which would as well be entirely incompatible with special
relativity \e{for such a particle}.)

Thus neither the Klein-Gordon equation nor the Dirac Hamiltonian are
capable of sensibly describing \e{strictly solitary-particle relativ%
istic quantum mechanics}~\ct{B-D}.  One might have thought that this
would have prompted the abandonment of those two constructs in favor
of the natural relativistic free-particle Hamiltonian $\sqrt{m^2c^4 +
|c\qvp|^2}$ of Eq.~(1), which is positive definite and has no problem
whatsoever with sensibly describing free solitary-particle relativistic
quantum mechanics.  Quite to the \e{contrary}, however, the Klein-Gordon
and Dirac wave functions have been duly quantized as field operators,
with the Hermitian conjugates of the \e{negative-energy parts} of those
quantum fields \e{reinterpreted} as \e{antiparticle} quantum fields%
~\ct{B-D}.  While antiparticles are an unquestionable feature of the
physical landscape, it is as well unquestionable that antiparticles are
\e{fully distinguishable} from their particle partners, a quality that
would \e{normally require} any one of them to be described by a quantum
field which is \e{completely independent} of the field that describes
its particle partner.  Furthermore, the underlying CP symmetry between
particles and their antiparticle partners is empirically known to be very
slightly \e{broken} (this certainly \e{doesn't} conflict with common sense
in view of the notable excess of particles over antiparticles in the world
around us!).  Such a breaking is well-nigh \e{impossible to achieve
theoretically} if the particle \e{and} its antiparticle partner \e{both}
``spring'' from the \e{very same quantum field} that was \e{originally
constructed} to describe \e{only the particle}.  However, CP symmetry
breaking is theoretically achievable in \e{myriad} ways \e{if} the
particle and its antiparticle partner are \e{each} described by its
\e{own independent} quantum field (as one elementary example, it can
be effected by introducing a tiny mass difference between those two
\e{independent} fields).  

In light of the above considerations, it is by \e{no} means apparent that
the \e{negative energy} solutions of the Klein-Gordon and Dirac
theories, which clearly \e{preclude} them from making sense in the
context of \e{strictly solitary-particle} relativistic quantum mechanics,
\e{are} in fact the ``triumph'' for the understanding of antiparticles
that they are conventionally claimed to be~\ct{B-D}.  To the contrary,
it is a field-theoretic \e{anomaly} that the \e{fully distinguishable}
antiparticle \e{fails} to be described by a field which is \e{completely
independent} of the field that describes its particle partner, and the
\e{need} for such \e{field independence} becomes \e{pressing} in view of
the empirically verified \e{breaking} of particle-antiparticle CP symme%
try.  Now the problematic \e{negative-energy} solutions of the Klein-%
Gordon and Dirac theories \e{only} arose as \e{entirely extraneous
artifacts} of either the \e{physically unmotivated iteration} or the like%
wise \e{physically unmotivated linearization} of the \e{classical-physics-%
mandated} natural relativistic free-particle quantum mechanics Hamiltonian
$\sqrt{m^2c^4 + |c\qvp|^2}$, which is \e{itself entirely positive}.
Given the \e{failure} of those negative-energy solutions to \e{properly
fulfill} their envisaged antiparticle role, which comes \e{in addition}
to their \e{extraordinarily artificial origin}, it is reasonable to begin
exploring possible alternatives.  The one that, of course, commands our
full attention is the \e{unconditional return} to the unexceptionable soli%
tary-particle \e{quantum mechanics} implied by Eq.~(1), simply because free
solitary-particle \e{classical} relativistic dynamics \e{mandates} for quan%
tum mechanics its Hamiltonian
$\sqrt{m^2c^4 + |c\qvp|^2}$!  Any disrespect of the \e{classical correspon%
dence principle}, even in \e{very small details}, is \e{highly likely} to
generate \e{wrong quantum mechanics}.  This \e{extremely tight coupling} of
quantum to classical mechanics is \e{implicit} in the Hamiltonian path
integral, which generates \e{all} quantum transition amplitudes \e{directly}
from the \e{purely classical Hamiltonian function}!  Now the Hamiltonian
path integral \e{hadn't been formulated} at the time that Klein, Gordon,
Schr\"{o}dinger and Dirac were \e{taking liberties} with Eq.~(1),
so the classical correspondence principle was \e{a vastly
less constraining concept} in the minds of these pioneers than what the
\e{fully mature theoretical underpinning} of quantum mechanics in fact
\e{implies}.  Even allowing for this, looking at some of the
\e{predictions} of the \e{classically-mandated relativistic quantum
mechanics Hamiltonian}
$\sqrt{m^2c^4 + |c\qvp|^2}$ for the free relativistic
\e{solitary particle} versus those
of Dirac's \e{linearized version of it}, i.e., $\qH_D$, is
\e{starkly revealing}: the former's \e{particle speed operator}
is $c|\qvp|/\sqrt{|\qvp|^2 + m^2c^2}$, which is \e{strictly less} than
$c$, while the latter's is simply $\sqrt{3}c$, a \e{universal speed} that
exceeds $c$ by over 70\%; the \e{lower bound} of the former's energy is
$mc^2$, while the latter's energies are negatively unbounded; the former's
acceleration operator \e{vanishes identically}, in accord with Newton's
first law for a free particle, while the latter has a \e{minimum} acceler%
ation magnitude of the order of the ``Compton acceleration'' $mc^3/\h$,
which for the electron works out to approximately $10^{28}g$; the
nonrelativistic limit of the former is \e{unambiguous} and \e{correct}, i.e.,
$(\sqrt{m^2c^4 + |c\qvp|^2} - mc^2)\rta |\qvp|^2/(2m)$ as $c\rta\infty$,
a result \e{thwarted} by the negative energies of the latter; the former
\e{conserves} particle orbital angular momentum, $\qvL = \qvx\times\qvp$,
which a \e{free particle} of \e{any spin} must do, the latter does not.
Dirac \e{specifically made certain that} $(\qH_D)^2 = m^2c^4 + |c\qvp|^2$,
which is the \e{same} as $(\sqrt{m^2c^4 + |c\qvp|^2}\:)^2$, but the
\e{tolerance} of quantum mechanics for \e{any alterations} of the character
of its classical input can be \e{poor in the extreme}!

If we now \e{accept} the positive definite Hamiltonian $\sqrt{m^2c^4 +
|c\qvp|^2}$ as the correct description of the \e{quantum mechanics} of
any free relativistic nonzero-mass solitary particle, then it obviously
\e{must} similarly apply to any free solitary \e{antiparticle}, albeit,
of course, with \e{that antiparticle's} degrees of freedom.  Since
particle and antiparticle are \e{fully distinguishable}, their wave-%
function \e{second quantization} will, in the \e{absence} of any
interaction, involve \e{two completely independent quantum fields},
whose operator evolutions are \e{both} determined by this \e{type} of
first-quantized Hamiltonian.  The pair-grouping of antiparticles with
their particle partners is then the result of an overall field theoretic
(slightly) \e{broken} CP symmetry, and is analogous to particle groupings
into isospin multiplets, SU(3) octets and \e{all the other particle
groupings that are the result of broken symmetries}!  With regard to the
``conventional'' approach to antiparticles, it seems implausible theoret%
ical physics to \e{specifically reject} the \e{classically-mandated} rel%
ativistic free solitary-particle quantum mechanics Hamiltonian
$\sqrt{m^2c^4 + |c\qvp|^2}$, that has \e{positive definite} energy,
simply in order to \e{make use} of its \e{physically unmotivated and
apparently defective Klein-Gordon and Dirac offshoots} for the express
purpose of \e{inventing from whole cloth} a scheme of ``partial-quantum-%
field negative-energy reinterpretation'' that \e{only applies} to the
pair-grouping of antiparticles with their particle partners amongst
\e{all the similar} particle-grouping schemes that contrariwise are due
to \e{broken symmetries}, and that appears to be \e{not capable of
coexisting} with the circumstantially expected and empirically verified
\e{breaking} of the underlying CP symmetry.

Whereas the Hamiltonian $\sqrt{m^2c^4 + |c\qvp|^2}$ provides a physically
sensible description of the relativistic quantum mechanics of a \e{free}
nonzero-mass solitary particle (and also of such a \e{free} solitary anti%
particle), the \e{focus} of this article lies \e{beyond} the merely
\e{free} relativistic solitary particle: it is on working out
Hamiltonians which describe physically sensible relativistic quantum
mechanics for such a solitary particle (of either spin 0 or spin $\hf$) in
\e{interaction} with an external electromagnetic field.  First, however,
we need to learn how to get past the technical stumbling block that
\e{neither} Hamiltonians \e{nor} the familiar \e{usual form} of the
Schr\"{o}dinger equation are \e{themselves} manifestly Lorentz covariant.

\subsection*{The four-momentum method in relativistic solitary-particle 
 mechanics}

When a solitary particle of nonzero mass $m$ \e{interacts} with
external fields, its relativistic Hamiltonian is of the general
form $H(\vx , \vP , t;m)$, where $\vx$ is the vector of the
particle's three position coordinates and $\vP$ is its \e{total
dynamical three-momentum}, i.e., the \e{sum} of its \e{kinetic}
three-momentum $\vp$ with any three-momentum contributions that
arise from its \e{interaction} with the external fields.  The
reason for the occurrence of the particle's \e{total} dynamical
three-momentum $\vP$ in its Hamiltonian is that the \e{total}
three-momentum of \e{any} physical system \e{generates} the
\e{translations} of that system's center-of-mass coordinates
$\vx_{\mr{CM}}$.  For the solitary particle, obviously $\vx_{\mr{CM}}
= \vx$, and the fact that $\vP$ \e{generates} the \e{translations} of
$\vx$ implies that the three components of $\vP$ are \e{canonically
conjugate} to the three corresponding components of $\vx$.  Therefore
it is the the solitary particle's \e{total} dynamical three-momentum $\vP$
that \e{properly belongs} in its \e{Hamiltonian} $H(\vx , \vP , t;m)$.
Furthermore, the \e{total dynamical four-momentum} of the solitary
particle is obviously $P^{\mu}\eqdf (H(\vx , \vP , t;m)/c,\vP )$.
Special relativity of course \e{imposes} on $P^{\mu}$ the
\e{requirement} that it transform between inertial frames as a
\e{Lorentz-covariant four-vector}.

The inherently \e{four-momentum
character} of solitary-particle relativistic dynamics naturally carries
over to its \e{quantum mechanics}.  The quantum expression of the
canonically conjugate character of $\vx$ to $\vP$ is, of course, the
familiar commutation relation, $[(\qvx)^i,(\qvP)^j]=i\h\dl_{ij}\id$,
which, in configuration representation, implies the familiar relation,
$\la\vx|\qvP|\psi(t)\ra=-i\h\del_{\vx}\la\vx|\psi(t)\ra$.  This, when
combined with the relativistic solitary-particle Schr\"{o}dinger equation,
$i\h\partial\la\vx|\psi(t)\ra/\partial t=\la\vx|\qH|\psi(t)\ra$, produces
the formal quantum-mechanical \e{equality of two four-vectors},
\[i\h\partial\la\vx|\psi(t)\ra/\partial x_{\mu}=\la\vx|\qP^{\mu}|\psi(t)\ra,\]
where the \e{covariant components} of $x_{\mu}=(ct,-\vx)$.
It is interesting to note that \e{iteration}
of this \e{four-momentum} Schr\"{o}dinger equation, followed by \e{index
contraction}, produces the Lorentz-invariant \e{generalization of the
Klein-Gordon equation}, $\pa^{\mu}\pa_{\mu}\la\vx|\psi(t)\ra =
-\la\vx|\qP^{\mu}\qP_{\mu}|\psi(t)\ra/\h^2$.  Of course this Lorentz
\e{scalar} equation can \e{not} be expected to \e{imply} the above Lorentz
\e{four-vector} Schr\"{o}dinger equation, and the \e{iteration} which
is part of its \e{derivation} can be expected to \e{burden} it with
\e{extraneous, unphysical} solutions (at least in the
strictly solitary-particle regime), such as the \e{negative-energy} ones
which have been previously noted above in the \e{free-particle} situation
that $\qP^{\mu}=\qp^{\mu}=(\sqrt{m^2c^2 + |\qvp|^2},\qvp)$.

The relativistic \e{free} particle is, of course, a \e{special case} of
the relativistic solitary particle, whose \e{total} dynamical three-momentum
$\vP$ consists of \e{only} its \e{kinetic} three-momentum $\vp$, which has
a value that depends on the choice of inertial frame from which it
is viewed in accord with the rules of special relativity for a free
particle of nonzero mass $m$.  Thus when viewed from the free particle's
rest frame, $\vp =\vz$, whereas when viewed from an inertial frame
in which the free particle's rest frame has velocity $\vv$, the free
particle's kinetic momentum $\vp$ is equal to $m\vv\gm$, where the Lorentz
time dilation factor $\gm\eqdf 1/\sqrt{1-|\vv /c|^2}$.  We therefore
calculate that to Lorentz boost the free particle from rest to kinetic
three-momentum $\vp$ involves the Lorentz time dilation factor $\gm (\vp )=
\sqrt{1 + |\vp /(mc)|^2}$ and the Lorentz boost velocity $\vv (\vp )=
c\vp /\sqrt{m^2c^2 + |\vp |^2}$.  We recall from the preceding paragraph
that the free particle's \e{dynamical} four-momentum $p^{\mu}\eqdf (\Hfr
(\vx , \vp , t;m)/c, \vp )$ \e{must} Lorentz transform between these two
inertial frames as a covariant four-vector.  In the free particle's rest
frame, $p^{\mu} = (\Hfr (\vx , \vz , t;m)/c, \vz )$, which has \e{only} its
nought component nonzero.  Therefore, it can be boosted to the inertial frame
where the free particle has kinetic three-momentum $\vp$ by using \e{only
the following four entries} of the sixteen-entry Lorentz boost matrix,
\[\Lambda^{\mu}_0(\vv (\vp )) = (\gm (\vp ), \gm (\vp )\vv (\vp )/c) =
(\sqrt{1 + |\vp /(mc)|^2}, \vp /(mc)).\]
This produces the \e{boosted} free-particle four-momentum,
\[ (\Hfr (\vx , \vz , t;m)\sqrt{1 + |\vp /(mc)|^2}/c,
\Hfr (\vx , \vz , t;m)\vp /(mc^2)),\] which is, of course,
\e{required} by the above-stated \e{imposition of Lorentz covariance} to be
\e{equal to},\[ p^{\mu}\eqdf (\Hfr (\vx , \vp , t;m)/c, \vp ).\]
Therefore Lorentz covariance of the free-particle four-momentum implies
that $\Hfr (\vx , \vz , t;m) = mc^2$, and, furthermore, that,
\[\Hfr (\vx , \vp , t;m) = \sqrt{m^2c^4 + |c\vp |^2}.\]
Thus we see that the imposition of special relativity
\e{completely determines} the Hamiltonian $\Hfr (\vx , \vp , t;m)$ of
\e{any} free particle of nonzero mass $m$ to be $\sqrt{m^2c^4 + |c\vp |^2}$.
This \e{uniqueness} of the relativistic $\Hfr$ in \e{classical}
relativistic dynamics obviously \e{also} enforces Eq.~(1) as the correct
\e{quantum mechanical} description of a \e{free} relativistic nonzero-mass
solitary particle.

For completeness we point out that $\Hfr$ can also be worked out
from the extraordinarily simple-looking Lorentz-invariant action
for the solitary free particle,
\[\int d\tau\,(-mc^2),\]
where $d\tau$ is the solitary particle's differential proper time
interval, which is defined via the particle's space-time contravariant
four-vector location $x^{\mu}=(ct, \vx)$ and,
\[(d\tau)^2\eqdf dx^{\mu}dx_{\mu}/c^2=(dt)^2-|d\vx/c|^2.\]
Therefore the relativistic Lagrangian $L$ for the solitary free
particle follows from its above Lorentz-invariant action as,
\[L=(-mc^2)d\tau/dt=(-mc^2)\sqrt{1-|\dot{\vx}/c|^2}=(-mc^2)/\gm,\]
where $\gm\eqdf 1/\sqrt{1-|\dot{\vx}/c|^2}$ is the usual Lorentz
time-dilation factor.  With the Lagrangian $L$ in hand, we can
work out the free-particle canonical momentum in the usual way,
\[\vp=\del_{\dot{\vx}}L=m\dot{\vx}\gm.\]
Continuing along these lines in classical dynamics textbook
fashion permits us to eventually calculate the free-particle
Hamiltonian $\Hfr$, but the process is astonishingly
long-winded, in contrast with the stark simplicity of the
above Lorentz-invariant action and its Lorentz-invariant
``time-dilated Lagrangian'', $-mc^2$.

\subsection*{Development of relativistic Hamiltonians for interacting
solitary particles}

Knowing that \e{any} solitary free particle of nonzero mass $m$
is described by the familiar relativistic square-root Hamiltonian,
$\qHfr =\sqrt{m^2c^4 + |c\qvp |^2}$ and four-momentum $\q{p}^{\mu}
=(\qHfr, \qvp)$, we now turn to the development of relativistic
Hamiltonians for such a solitary particle in \e{interaction} with
external fields.  For an external \e{electromagnetic} field
there do exist \e{nonrelativistic Hamiltonians} for the spinless
and spin~$\hf$ charged particle which are considered \e{trustworthy},
and it is physically clear that for a solitary particle \e{the
correct nonrelativistic Hamiltonian ought to actually determine
its fully relativistic counterpart}!  That is because the
\e{correct nonrelativistic Hamiltonian} ought to be \e{absolutely
precise in the instantaneous rest frame of the particle}.
This line of physical reasoning seems more a route toward relativistic
upgrading of the dynamics itself rather than the Hamiltonian
directly, albeit the latter can presumably always be extracted
from the former.

A direct relativistic upgrade of \e{any term} of a nonrelativistic
Hamiltonian would involve crafting a \e{properly Lorentz-covariant
four-momentum} whose nought component times $c$ \e{reduces} to
that particular term of the nonrelativistic Hamiltonian in
the particle's rest frame.  If this is done for \e{all the terms}
which make up the nonrelativistic Hamiltonian, then all
those four-momenta are to be added together.  As was discussed
in the previous section, the resulting \e{total three-momentum} will
be canonically conjugate to the three components of the particle's
position vector.  The resulting \e{total energy} therefore becomes
the interacting particle's \e{Hamiltonian} if its dependence on
the particle's \e{kinetic} three-momentum can be reexpressed as
dependence on its \e{total} three-momentum, i.e., we need to
\e{solve} for the \e{kinetic} three-momentum as a \e{function}
of the \e{total} three-momentum and then \e{substitute} this
function into the total energy, which thereupon becomes the
particle's relativistic \e{Hamiltonian}.  It cannot be guaranteed,
of course, that the particle's kinetic momentum can be obtained
as a function of its total momentum in \e{closed form}.  If
this function cannot be obtained in closed form, successive
iteration approximations to it are sometimes adequate.

Before turning to our first example of this approach, we
must note a slight but universal exception to the rule
that the nought component times $c$ of each of our Lorentz
covariant four-momenta reduces in the particle's rest frame
to a corresponding term of the nonrelativistic Hamiltonian.
The nonrelativistic single-particle Hamiltonian will, of
course, always have a kinetic energy term, whose
nonrelativistic form vanishes by convention in the particle's
rest frame, but whose \e{relativistic counterpart} always
tends toward $mc^2$ in the particle's rest frame, where $m$
is the particle's rest mass.  Indeed we are well aware that
the Lorentz covariant four-momentum which corresponds to the
nonrelativistic single-particle Hamiltonian's kinetic energy
term is simply the \e{universal relativistic free-particle
four-momentum} $p^{\mu}$, which, as we pointed out in the
previous section, \e{has the immutable square-root form},
$p^{\mu}=(\sqrt{m^2c^2 + |\vp |^2}, \vp )$, where $\vp$
is the particle's \e{kinetic three-momentum}.

\subsection*{Relativistic Hamiltonian for the spinless charged
solitary particle}

A completely nonrelativistic spinless charged particle which
interacts with an external electromagnetic field is described
by the Hamiltonian,
\re{
\Hnz = |\vp |^2/(2m) + eA^0(\vx ,t). 
}{2}
Now in order to take account of the interaction of a
nonrelativistic spinless charged particle with an external
\e{magnetic} field, the Hamiltonian of Eq.~(2) is often
upgraded to a \e{partially relativistic form} which involves
the electromagnetic vector potential $\vA $ occurring
in conjunction with the speed of light $c$ in a term of
the form $e\vA (\vx ,t)/c$.  Such terms of course \e{vanish}
in the limit that $c\rta\infty$, which is why we have \e{left
them out} of the \e{completely nonrelativistic} Hamiltonian
of Eq.~(2): we shall see that they are a \e{natural
consequence} of the full relativistic upgrade of the
Hamiltonian of Eq.~(2) that we are about to undertake.

As we have discussed at the end of the previous section,
the kinetic energy term $|\vp |^2/(2m)$ of the nonrelativistic
Hamiltonian $\Hnz$ of Eq.~(2) corresponds
to the free-particle four-momentum,
$p^{\mu}=(\sqrt{m^2c^2 + |\vp |^2}, \vp )$.  The potential
energy term $eA^0(\vx ,t)$ of this nonrelativistic Hamiltonian
involves the nought component $A^0$ of the full electromagnetic
four-potential $A^{\mu}$.  Therefore the Lorentz covariant
four-momentum which corresponds to the potential energy
term $eA^0(\vx ,t)$ of this nonrelativistic Hamiltonian is,
$eA^{\mu}(\vx ,t)/c$.  Thus the \e{total} four-momentum
$P^{\mu}$ of our relativistically upgraded system is,
\[P^{\mu} = p^{\mu} + eA^{\mu}(\vx ,t)/c.\]  From this we read
off the total three-momentum of our relativistic system,
\[\vP = \vp + e\vA (\vx ,t)/c,\] and its total energy,
\[E(\vx ,\vp ,t) = \sqrt{m^2c^4 + |c\vp |^2} +eA^0(\vx ,t).\]
We now recall from the previous section that in order to
obtain the relativistic Hamiltonian $H(\vx ,\vP ,t)$ we must
solve for the kinetic three-momentum $\vp$ as a function of
the total three-momentum $\vP$, i.e., we must solve for
$\vp (\vP )$, which when put into the total energy
$E(\vx ,\vp ,t)$, yields the relativistic Hamiltonian
$H(\vx ,\vP ,t)$ as, \[H(\vx ,\vP ,t)=E(\vx ,\vp (\vP ),t).\]
Fortunately, in this instance we are easily able to obtain
$\vp (\vP )$ in closed form, namely,
$\vp (\vP )= \vP -e\vA (\vx ,t)/c$.
Therefore we obtain the full relativistic upgrade of the
Hamiltonian of the spinless charged particle in interaction
with an external electromagnetic four-potential,
\re{
\Hrz = \sqrt{m^2c^4 + |c\vP -e\vA (\vx ,t)|^2} +
eA^0(\vx ,t).
}{3}
It is clear that as $c\rta\infty$, $(\Hrz - mc^2)\rta\Hnz$.  We also note
that since the argument of the square root in Eq.~(3) is a joint function
of $\vP$ and $\vx$, the \e{operator ordering ambiguity} which is a
consequence of Dirac's \e{original} widely accepted canonical commutation
rules has the potential to present a considerable annoyance for the
quantization of $\Hrz$.  Fortunately, however, \e{both} the Hamiltonian
path integral~\ct{K-S} \e{and} a self-consistent slight strengthening of
Dirac's original canonical commutation rules~\ct{Ka} have been shown to
yield \e{exactly the same} completely \e{unambiguous} Born-Jordan quanti%
zation of \e{all} classical dynamical variables: this aspect of quantiza%
tion would appear to be of considerably greater practical relevance to
\e{relativistic} quantum mechanics than to its nonrelativistic counterpart.
It may be of interest to the reader that Dirac's well-known but inadequate
phase-space-vector Cartesian-component canonical commutation rules are to
be replaced by the slightly stronger, \e{but still self-consistent},
canonical commutation rule~\ct{Ka},
\[ [f_1(\qvx)+g_1(\qvP),f_2(\qvx)+g_2(\qvP)]=i\hbar
(\,\Q{(\del_{\vx}f_1(\vx)\dt\del_{\vP}g_2(\vP))}
-\Q{(\del_{\vP}g_1(\vP)\dt\del_{\vx}f_2(\vx))}\,),\]
where \e{both the hat accent and the overbrace} are used to
indicate the \e{quantization of a classical dynamical variable}.

It is furthermore apparent that the \e{classical} Hamiltonian equations
of motion which the $\Hrz$ of Eq.~(3) \e{implies} are simply those of the
very well-known \e{fully relativistic Lorentz force}~\ct{L-L}.  For
completeness we also mention that the corresponding Lorentz-invariant
action is the quite well-known,
\[\int d\tau\,(-mc^2-(e/c)A_{\mu}(x^{\nu})dx^{\mu}/d\tau),\]
and from it $\Hrz$ can as well be calculated by a somewhat tedious
standard sequence of classical dynamics steps.  The pattern that emerges
here for the solitary spinless particle in fully relativistic interaction
with an external electromagnetic field is a \e{precise relationship} to
\e{well-known} classical relativistic dynamics which the Klein-Gordon
equation cannot even \e{begin} to achieve.  This \e{precise classical rel%
ativistic correspondence} lends impressive support to the quantization of
$\Hrz$ being the \e{correct} quantum mechanics description of a relativis%
tic spinless nonzero-mass charged solitary particle in interaction with an
external electromagnetic field, and validates the \e{methodology} whereby
the relativistic Eq.~(3) was derived from the nonrelativistic Eq.~(2).

\subsection*{Relativistic Hamiltonian for the spin~$\hf$ charged
solitary particle} 

We now turn to the relativistic upgrade of the nonrelativistic
Pauli Hamiltonian for the interaction of a charged particle of
spin~$\hf$ and specified magnetic moment $g$-factor with an
external electromagnetic four-potential $(A^0, \vA )$,
\re{
\Hpl = \Hnh =|\vp -e\vA (\vx ,t)/c|^2/(2m) + eA^0(\vx ,t) +
(ge/(mc))(\h /2)\vec\sg\dt(\del_{\vx}\times\vA (\vx ,t)).
}{4}
There are some points to bear in mind about $\Hnh$.  First it is
\e{already} partially relativistic.  Note that if we take the
limit $c\rta\infty$, $\Hnh\rta\Hnz$, i.e., we \e{lose} the
electromagnetic field's interaction with spin~$\hf$ just as surely
in the the fully nonrelativistic limit as we lose it in the
the $\h\rta 0$ \e{classical} limit.  It is apparent that $\Hnh$
\e{must retain} relativistic effects through order $O(1/c)$ in
order to be able to describe the magnetic moment coupling of
spin~$\hf$.  Second, $\Hnh$ is a Hermitian two-by-two \e{matrix}.
Therefore the relativistic four-momenta we intend to develop
must be such matrices as well.  That should not be a problem
so long as we \e{do not generate} four-momentum components that
\e{fail to mutually commute}.  Therefore it would
be prophylactic to \e{quarantine} the \e{one} intrinsically two-%
by-two matrix term of $\Hnh$, i.e., the spin-$\hf$/magnetic-field
interaction, $(ge/(mc))(\h /2)\vec\sg\dt(\del_{\vx}\times\vA (\vx ,t))$,
\e{within a Lorentz invariant}.  This, however, raises the issue that
we wish such nonrelativistic Hamiltonian terms to correspond to $c$
times the nought component of a \e{four-momentum} in the particle's
rest frame, \e{not} to a Lorentz \e{invariant}.  But that turns out
to be quite easily arranged once the Lorentz invariant is in hand:
we simply \e{divide} that Lorentz-invariant spin interaction energy
by $mc^2$, and then multiply the resulting dimensionless
Lorentz-invariant object by the \e{free-particle} four-momentum
$p^{\mu} = (\sqrt{m^2c^2 + |\vp |^2},\vp )$, where $\vp$ is, of
course, the particle's kinetic three-momentum.

From our work in the previous section with $\Hrz$, we recognize
that the ``partially relativistic'' kinetic energy term of
$\Hnh$, namely, $|\vp -e\vA (\vx ,t)/c|^2/(2m)$, \e{still} simply
corresponds to the \e{usual} free-particle four-momentum, namely
$p^{\mu} = (\sqrt{m^2c^2 + |\vp |^2},\vp )$.  We of course
immediately as well recognize that the term $eA^0(\vx ,t)$ of
$\Hnh$ corresponds to the four-momentum, $eA^{\mu}(\vx ,t)/c$.

Now comes the difficult part: we are to quarantine the one
intrinsically two-by-two matrix term of $\Hnh$, namely,
$(ge/(mc))(\h /2)\vec\sg\dt(\del_{\vx}\times\vA (\vx ,t))$, within
a Lorentz invariant.  To move toward that goal, we note that the
Lorentz \e{covariant} form of the electromagnetic field is the
second-rank antisymmetric tensor $F^{\mu\nu} = \partial^{\mu}A^{\nu} -
\partial^{\nu}A^{\mu}$.  Now we can write a magnetic-field axial
vector component from $\Hnh$, namely, $(\del_{\vx}\times\vA )^i$, as,
\[ (\del_{\vx}\times\vA )^i = \ep^{ijk}\partial^jA^k =
\hf\ep^{ijk}(\partial^jA^k - \partial^kA^j) =\hf\ep^{ijk}F^{jk}.\]
Therefore we obtain that, \[ (\h /2)\vec\sg\dt(\del_{\vx}\times\vA ) =
\hf (\h /2)\sg^i\ep^{ijk}F^{jk} = \hf (\h /2)\ep^{jki}\sg^iF^{jk}.\]
We now \e{define}, for the particle in its rest frame, the second-rank
antisymmetric spin tensor, $S^{jk}\eqdf (\h /2)\ep^{jki}\sg^i$.  Then
$(\h /2)\vec\sg\dt(\del_{\vx}\times\vA ) = \hf S^{jk}F^{jk}$.  To reach
our final goal, we must define a Lorentz covariant $s^{\mu\nu}$ such
that $s^{\mu\nu}F_{\mu\nu}$ reduces to $S^{jk}F^{jk}$ in the particle's
rest frame.  We need \e{only} specify $s^{\mu\nu}$ in a \e{single}
inertial frame for it to be uniquely defined: it is then obtained in
\e{any} inertial frame via the appropriate Lorentz transformation.
So let us simply specify $s^{\mu\nu}$ in the particle's rest frame
as follows: $s^{00} = 0$, $s^{i0} = s^{0i} = 0$, and $s^{ij}=S^{ij} =
(\h/2)\ep^{ijk}\sg^k$, $i,j = 1,2,3$.  With that definition, we do
indeed have that $s^{\mu\nu}F_{\mu\nu}$ reduces to $S^{jk}F^{jk}$ in
the particle's rest frame.  To obtain $s^{\mu\nu}$ in the frame where
the particle has kinetic three-momentum $\vp$, we must apply the
appropriate Lorentz boost to its two indices,
\re{
s^{\mu\nu}(\vp ) = \Lambda^{\mu}_i(\vv (\vp ))\Lambda^{\nu}_j(\vv (\vp ))
(\h /2)\ep^{ijk}\sg^k,
}{5}
where, of course, the Lorentz boost velocity $\vv (\vp ) =
c\vp /\sqrt{m^2c^2 + |\vp |^2}$ and the Lorentz time dilation
factor $\gm(\vp ) = \sqrt{1 + |\vp /(mc)|^2}$.  We note that the
antisymmetry of $\ep^{ijk}$ in its two indices $i$ and $j$ implies
that $s^{\mu\nu}(\vp )$ is an antisymmetric tensor in its two
indices $\mu$ and $\nu$.

Thus the two-by-two matrix spin~$\hf$ interaction term of $\Hnh$,
namely $(ge/(mc))(\h /2)\vec\sg\dt(\del_{\vx}\times\vA (\vx ,t))$,
is now safely quarantined as the Lorentz invariant
$(g/2)(e/(mc))s^{\mu\nu}(\vp )F_{\mu\nu}(\vx ,t)$.
We proceed to write down the corresponding four-momentum for this
term by following the instructions that were given above, namely to
divide this Lorentz invariant by $mc^2$ and then multiply the result
into the free-particle four-momentum $p^{\mu}$.  With that, we are
now in a position to write down the \e{total} four-momentum,
\[ P^{\mu} =
p^{\mu} (1 + (g/2)(e/(m^2c^3))s^{\al\bt}(\vp )F_{\al\bt}(\vx ,t)) +
eA^{\mu}(\vx ,t)/c,\]
from which we obtain the total energy,
\re{
E(\vx ,\vp ,t) = \sqrt{m^2c^4 +|c\vp |^2}
(1 + (g/2)(e/(m^2c^3))s^{\mu\nu}(\vp )F_{\mu\nu}(\vx ,t))
+ eA^0(\vx ,t),
}{6}
and the total three-momentum,
\re{
\vP = \vp (1 + (g/2)(e/(m^2c^3))s^{\mu\nu}(\vp )F_{\mu\nu}(\vx ,t))
+ e\vA (\vx ,t)/c.
}{7}
It is obvious from Eq.~(7) that we \e{cannot solve} for $\vp (\vP )$
in \e{closed form}, but we \e{can} write $\vp (\vP )$ in
``iteration-ready'' form as,
\re{
\vp (\vP ) = (\vP - e\vA (\vx ,t)/c)
(1 + (g/2)(e/(m^2c^3))s^{\mu\nu}(\vp (\vP ))F_{\mu\nu}(\vx ,t))^{-1},
}{8}
and, of course, from $E(\vx ,\vp (\vP ),t)$, we also obtain the
schematic form of the relativistic Hamiltonian,
\re{
\Hrh(\vx ,\vP ,t) =
\sqrt{m^2c^4 +|c\vp (\vP )|^2}
(1 + (g/2)(e/(m^2c^3))s^{\mu\nu}(\vp (\vP ))F_{\mu\nu}(\vx ,t))
+ eA^0(\vx ,t).
}{9}
If we take the limit $g\rta 0$ in Eqs.~(8) and (9), then
$\Hrh(\vx ,\vP ,t)\rta\Hrz(\vx ,\vP ,t)$, as is easily checked
from Eq.~(3).  Of course it is nothing more than the most
basic common sense that fully relativistic spin~$\hf$ theory
simply reduces to fully relativistic spinless theory when the
spin coupling of the single particle to the external field is
switched off, but analogous cross checking between the Dirac
and Klein-Gordon theories is never even discussed!  It may also
be checked that $(\Hrh - mc^2)$ agrees with the ``partially
relativistic'' Pauli $\Hnh$ through terms of order $O(1/c)$
in the limit $c\rta\infty$.

It is unfortunate that Eq.~(8) for $\vp (\vP )$ is not amenable
to closed-form solution, but if we assume that the spin coupling term,
$(g/2)(e/(m^2c^3))s^{\mu\nu}(\vp (\vP ))F_{\mu\nu}(\vx ,t)$, which is
a dimensionless Hermitian two-by-two matrix, effectively has the
magnitudes of both of its eigenvalues much smaller than unity (which
should be a very safe assumption for atomic physics), then we can
approximate $\vp (\vP )$ via successive iterations of Eq.~(8), which
produces the approximation $(\vP - e\vA (\vx ,t)/c)$ for $\vp (\vP )$
through zeroth order in the spin coupling and,
\[\vp (\vP )\approx  (\vP - e\vA (\vx ,t)/c)
(1 + (g/2)(e/(m^2c^3))s^{\mu\nu}(\vP - e\vA (\vx ,t)/c)
F_{\mu\nu}(\vx ,t))^{-1},\]
through first order in the spin coupling.  We wish to
interject at this point that since $s^{\mu\nu}(\vp (\vP ))$
is an antisymmetric tensor, the tensor contraction
$s^{\mu\nu}(\vp (\vP ))F_{\mu\nu}(\vx ,t)$ is equal to
$2s^{\mu\nu}(\vp (\vP ))\partial_{\mu}A_{\nu}(\vx ,t)$,
which is often a more transparent form. Now if we simply
use the approximation $(\vP - e\vA (\vx ,t)/c)$
through zeroth order in the spin coupling for $\vp (\vP )$,
we obtain the following approximation to $\Hrh$,
\re{
\Hrh(\vx ,\vP ,t)\approx
\sqrt{m^2c^4 +|c\vP - e\vA (\vx ,t)|^2}
(1 + (ge/(m^2c^3))s^{\mu\nu}(\vP - e\vA (\vx ,t)/c)
\partial_{\mu}A_{\nu}(\vx ,t)) + eA^0(\vx ,t).
}{10}
For completeness we also point out the corresponding Lorentz-%
invariant action \e{might} plausibly be,
\[\int d\tau\,
(-mc^2(1+(g/2)(e/(m^2c^3))s^{\mu\nu}(dx^{\al}/d\tau)F_{\mu\nu}(x^{\bt}))-
(e/c)A_{\mu}(x^{\bt})dx^{\mu}/d\tau),\]
where $s^{\mu\nu}(c,\vz)$ in the particle's rest frame is specified
as follows: $s^{00}(c,\vz) = 0$, $s^{i0}(c,\vz) = s^{0i}(c,\vz) = 0$,
and $s^{ij}(c,\vz) = (\h/2)\ep^{ijk}\sg^k$, $i,j = 1,2,3$.  If the
above \e{guess} for the Lorentz-invariant action is correct, then in
principle $\Hrh$ can be tediously worked out from it, but we assuredly
\e{recommend against} trying to proceed by that route.  It is very clear
indeed that the Lorentz-covariant four-momentum approach keeps one in
vastly better contact with crucial physics than does trying to \e{guess}
the invariant action.  Furthermore, even a \e{correct guess} of that
action is separated by what is always a \e{very} considerable and tedious
calculational distance from the desired Hamiltonian. The four-momentum
method \e{seems in all respects} the best approach to \e{upgrading}
nonrelativistic solitary-particle Hamiltonians to relativistic ones.

With $\vA (\vx ,t) = \vz$ and $A^0(\vx ,t) = -e/|\vx |$, relativistic
and spin corrections to the nonrelativistic hydrogen atom energy
spectrum can be investigated by regarding $(\q\Hrh - mc^2 - \q\Hnz)$
as a perturbation of the familiar spinless nonrelativistic
Hamiltonian $\q\Hnz$, whose exact bound state solutions are well-known,
and to which $(\q\Hrh - mc^2)$ clearly reduces as $c\rta\infty$. The
effect of spin on the hydrogen energy spectrum arises from the fact
that $s^{i0}(\vp (\vP ))$ doesn't vanish if $\vp (\vP )$ is \e{nonzero},
i.e., a \e{moving} spin~$\hf$ particle has \e{spin coupling} to an
external \e{electric} field.  Another way to see this is to realize
that a purely electric field in an inertial frame in which the
particle is \e{moving} gives rise to a \e{magnetic field} in the
particle's \e{rest frame}, which thus activates the
\e{nonrelativistic} Pauli spin coupling in that rest frame.

\subsection*{Conclusion}

The discussion just above is a \e{pertinent reminder} that our \e{fully
relativistic} single particle Hamiltonians $\Hrz$ and $\Hrh$ have been
\e{completely founded} on the premise that well-known \e{nonrelativis%
tic} single-particle electromagnetic interactions are \e{exact} in the
particle's rest frame.  Their \e{only additional input} was the require%
ment of \e{strict Lorentz covariance}.  It is thus \e{no accident} that,
for example, the classical Hamiltonian equations of motion which are
implied by $\Hrz$ are \e{precisely those of the fully relativistic
Lorentz force law for a single charged particle}.  The fully relativis%
tic single-particle electromagnetic Hamiltonians $\Hrz$ and $\Hrh$ thus
represent the \e{epitome of theoretical physics conservatism} in the realm
of an electromagnetically interacting relativistic solitary particle.


\begin{thebibliography}{9}
\bibitem{B-D}
J. D. Bjorken and S. D. Drell,
\e{Relativistic Quantum Mechanics}
(McGraw-Hill, New York, 1964).
\bibitem{K-S}
E. H. Kerner and W. G. Sutcliffe,
J.\ Math.\ Phys.\ \textbf{11},
391 (1970).
\bibitem{Ka}
S. K. Kauffmann,
arXiv:0908.3755 [quant-ph]
(2009).
\bibitem{L-L}
L. D. Landau and E. M. Lifshitz,
\e{The Classical Theory of Fields}
(Butterworth-Heinemann, Oxford, 1975).
\end{thebibliography}
\end{document}